\documentclass[%
 reprint,
showpacs,
 amsmath,amssymb,
]{revtex4-1}

\usepackage{graphicx}
\usepackage{dcolumn}
\usepackage{bm}

\begin{document}

\title{Feedback-free optical cavity with self-resonating mechanism}%

\author{Y. Uesugi}
\affiliation{AdSM Hiroshima University, 1-3-1 Kagamiyama, Higashi Hiroshima, Hiroshima 739-8530, Japan}
\author{Y. Hosaka}
\affiliation{Research Institute for Science and Engineering, Waseda University, 3-4-1 Okubo, Shinjuku-ku 169-8555, Japan}
\author{Y. Honda}
\affiliation{High Energy Accelerator Research Organization (KEK), 1-1 Oho, Tsukuba, Ibaraki 305-0801, Japan}
\author{A. Kosuge}
\affiliation{High Energy Accelerator Research Organization (KEK), 1-1 Oho, Tsukuba, Ibaraki 305-0801, Japan}
\author{K. Sakaue}
\affiliation{Research Institute for Science and Engineering, Waseda University, 3-4-1 Okubo, Shinjuku-ku 169-8555, Japan}
\author{T. Omori}
\affiliation{High Energy Accelerator Research Organization (KEK), 1-1 Oho, Tsukuba, Ibaraki 305-0801, Japan}
\author{T. Takahashi}
\email{tohru-takahashi@hiroshima-u.ac.jp}
\thanks{Corresponding author}
\affiliation{AdSM Hiroshima University, 1-3-1 Kagamiyama, Higashi Hiroshima, Hiroshima 739-8530, Japan}
\author{J. Urakawa}
\affiliation{High Energy Accelerator Research Organization (KEK), 1-1 Oho, Tsukuba, Ibaraki 305-0801, Japan}
\author{M. Washio}
\affiliation{Research Institute for Science and Engineering, Waseda University, 3-4-1 Okubo, Shinjuku-ku 169-8555, Japan}

\date{\today}
\begin{abstract}
We demonstrated the operation of a high finesse optical cavity without utilizing an active feedback system to stabilize the resonance.
The effective finesse, which is a finesse including the overall system performance, of the cavity was measured to be $394,000 \pm 10,000$, 
and the laser power stored in the cavity was $2.52 \pm 0.13$ kW, which is approximately 187,000 times 
greater than the incident power to the cavity. 
The stored power was stabilized with a fluctuation of $1.7 \%$, and we confirmed continuous cavity operation for more than two hours.
This result has the potential 
to trigger an innovative evolution for applications that use optical resonant cavities
 such as compact photon sources with  laser-Compton scattering or cavity enhanced absorption
spectroscopy.
\end{abstract}


\maketitle

\section{Introduction}

Optical resonant cavities have been developed and used in a wide area of optical science applications. 
One attractive use of a resonant cavity is its ability to achieve a high power 
laser wave by coherently stacking laser waves in the cavity.
The maximum laser power stored in an optical cavity to date is 670 kW, 
which was stored with a relatively low power enhancement factor of approximately 1000 \cite{MPQ}.   
An optical cavity can be utilized in the generation of photon beams by laser-Compton scattering,  
which provides photon beams of a wide energy range 
from X rays for material science and industrial/medical applications to 
high energy  gamma rays for particle physics. 
The optical cavity is a novel solution for constructing 
high peak- and high average-power laser systems for photon sources 
 without requiring large scale laser facilities, 
and optical cavities have been developed and demonstrated in several projects\cite{Akagi, LUCX, LAL}.
In these recent projects, the laser power enhancement and stored laser power 
in the optical cavity have reached an order of $10^4$ and $10$ kW, respectively.
Cavity enhanced absorption spectroscopy (CEAS) utilizes the high finesse of an optical cavity.
In CEAS, the cavity detects trace element absorbers    
via a change of the finesse of the cavity caused by the absorbers\cite{springer}.
It must be noted that  narrow laser oscillation linewidth is not
required for either laser-Compton photon sources or CEAS. 

 To obtain a high power enhancement factor for laser-Compton scattering and high detection sensitivity for CEAS, one needs high finesse optical cavities with high reflectivity mirrors, 
which requires precise control of the optical path in the cavity. 
 For example, the full width of half maximum (FWHM) of the resonance of the cavity in Ref.~\onlinecite{Akagi} 
was 260 pm with a finesse of 4040; thus, the optical path length had to be controlled 
with a precision much smaller than 260 pm.
In Ref.~\onlinecite{Akagi}, the path length has been controlled 
with a precision of 20 pm using a sophisticated feedback system \cite{FBRSI}. 
In general, the required precision for the optical path 
is inversely proportional to finesse. 
For example, for the effective finesse of $394,000$
that we report in this study, 
which is the finesse including the overall performance of the 
self-resonating mechanism discussed in section \ref{sec:eval}, 
the FWHM of the resonance is approximately  $2$ pm. Thus, 
the optical path length must be controlled to be sufficiently 
smaller than  $2$ pm.
This value may not be unreachable, 
but will remain as a technical challenge, 
particularly for operation in non-isolated environments such as in accelerator facilities.
Recent examples of high finesse optical resonant cavities can be found 
in Ref.~\onlinecite{Millo, PVLAS} in which the finesse of approximately  $800,000$ and $789,000$ were achieved 
respectively.
However, it should be noted that these high finesse were achieved by sophisticated control of the equipment in a quiet environment.

In this study, we introduce a feedback-free optical cavity
with a self-resonating mechanism that successfully operates with an effective finesse of $394,000$ 
and an average stored power of $2.52$ kW,  which is
$187,000$ times greater than the input power.
The concept of a self-resonating cavity has been introduced
in our previous work \cite{honda}
in which the cavity is a part of the laser oscillator
embedded in the laser oscillation scheme.    
In this study, we achieved stable oscillation for the first time 
with a high finesse cavity, which opens new possibilities for realizing high finesse optical cavities for 
applications such as laser-Compton scattering.
A similar scheme has been discussed previously; however, 
the previous studies were not aiming to achieve high finesse and high power storage in the cavity\cite{harvey,cheng}.

\section{Principle of the Self-resonating Mechanism}

In this system, electromagnetic waves 
that  
satisfy the resonance condition with the cavity, i.e., $ L^{cav} = n\lambda$/2, are
chosen as seeds of the laser oscillation, 
where $L^{cav} $, $\lambda$, and $n$ are the length of the cavity, the wavelength, and an integer, respectively.
The seeds are amplified in a laser medium and the laser oscillation starts spontaneously 
if the amplifier gain exceeds the loss during the circulation.
The laser  is in a multi-mode oscillation state, i.e., the waves that satisfy the resonance condition 
can contribute to the oscillation, while the range of the wavelength is determined by characteristics of the
optical components of the system. 
Without any feedback mechanism for stabilizing $L^{cav}$, 
the cavity length fluctuates due to external disturbances such as acoustic noise in the environment, 
which in turn leads to fluctuation of the laser wavelength resonating with the cavity. 
However, the laser oscillation can, in principle, be developed 
with new wavelengths chosen by the cavity.
When the cavity fluctuates and goes to an off-resonant condition, 
the intensity of the laser wave decreases exponentially with the time constant
depending on the finesse. In the case of this study, the time constant is 
$145$ $\mu$s, as will be described later. 
The laser oscillation tends to resume at a new frequency;  
however, its time scale depends on various conditions such as 
the pump rate, the lifetime of the excited state in the lasing media, losses in the optical loop, 
the strength of the external disturbances, and the mechanical vibration property
of the cavity. 
The behavior of the self-resonating mechanism is a result of such multiple 
conditions and can only be examined experimentally.

\section{Experimental Setup}

\begin{figure*}
\includegraphics[width=0.35\linewidth,angle=270]{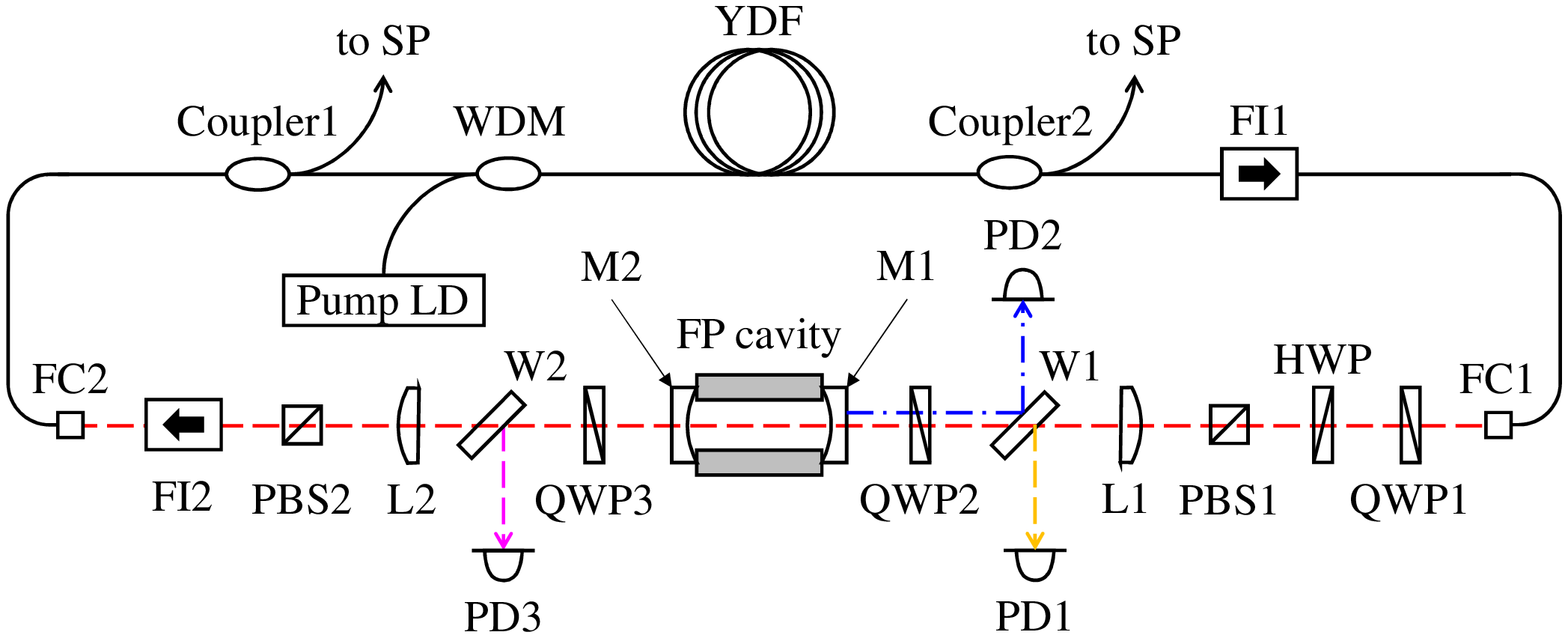}
\caption{The optical setup used to demonstrate the feedback-free optical cavity with the self-resonating mechanism.}
\label{fig:schematic}
\end{figure*}
 A schematic of the optical setup is shown in figure \ref{fig:schematic}. 
The optical system consists of an ytterbium-doped single mode fiber (YDF) amplifier,  a 
Fabry-Perot cavity, and other optical components to form the closed loop for laser oscillation.
All components are assembled on the optical table in air. 
The laser beam from the YDF  is released to free space via a collimator (FC1). 
The linear polarization in a specific direction is chosen by a polarized beam splitter (PBS1), 
while the polarization direction is adjusted with
 a set of quarter and half wave plates (QWP1 and HWP, respectively). 
The power of the laser beam incident to and reflected from 
the coupling mirror of the cavity, M1, is monitored by photodetectors (PD1 and PD2) 
by sampling part of the light with a wedge plate (W1). 
The transmitted light through the output mirror of the cavity, M2,
is coupled to the optical fiber via a fiber coupler (FC2). 
The transmitted light is sampled and monitored by a wedge plate (W2) and photodetector (PD3).
In addition to the two Faraday isolators (FI1 and FI2), which primarily determine the propagation direction,  
two sets of PBS and QWP (PBS1/QWP2 and PBS2/QWP3) are inserted to prevent unwanted stray light propagating 
backward to ensure stable operation of the system.
The two lenses, L1 and L2, are responsible for  optical coupling between  
the cavity and the fiber optics  
and for realizing  propagation of the fundamental transverse mode  throughout the loop, respectively.  
The YDF is pumped by a $330$-mW laser diode (LD) with a central wavelength of  978 nm
via a wavelength division multiplexer coupler (WDM).  
The length of the YDF is 38 cm, its  gain peaks at 1030 nm, and the overall gain, including
the optical losses in the system, peaks at 1047 nm.

The optical cavity consists of two high reflectivity mirrors, M1 and M2, mounted on a  super-invar alloy tube.
The distance between two mirrors, $L^{cav}$, is $208 \pm 1 $ mm, which corresponds to the free spectral range  (FSR) of 
$c/(2L^{cav}) = 721\pm 3$ MHz, where $c$ is the velocity of light in vacuum.
The mirrors, M1 and M2, are dielectric multilayer mirrors manufactured by ATFilms Co, 
whose reflectivity is  guaranteed to be more than 99.999$\%$ by the manufacturer.
Before implementation to the self-resonating loop, 
the finesse of the cavity was independently measured by the lifetime $\tau_c$ of the laser light in the cavity
with a wavelength of $1047$ nm (the cavity ring-down method).
\begin{figure}
\includegraphics[width=\linewidth]{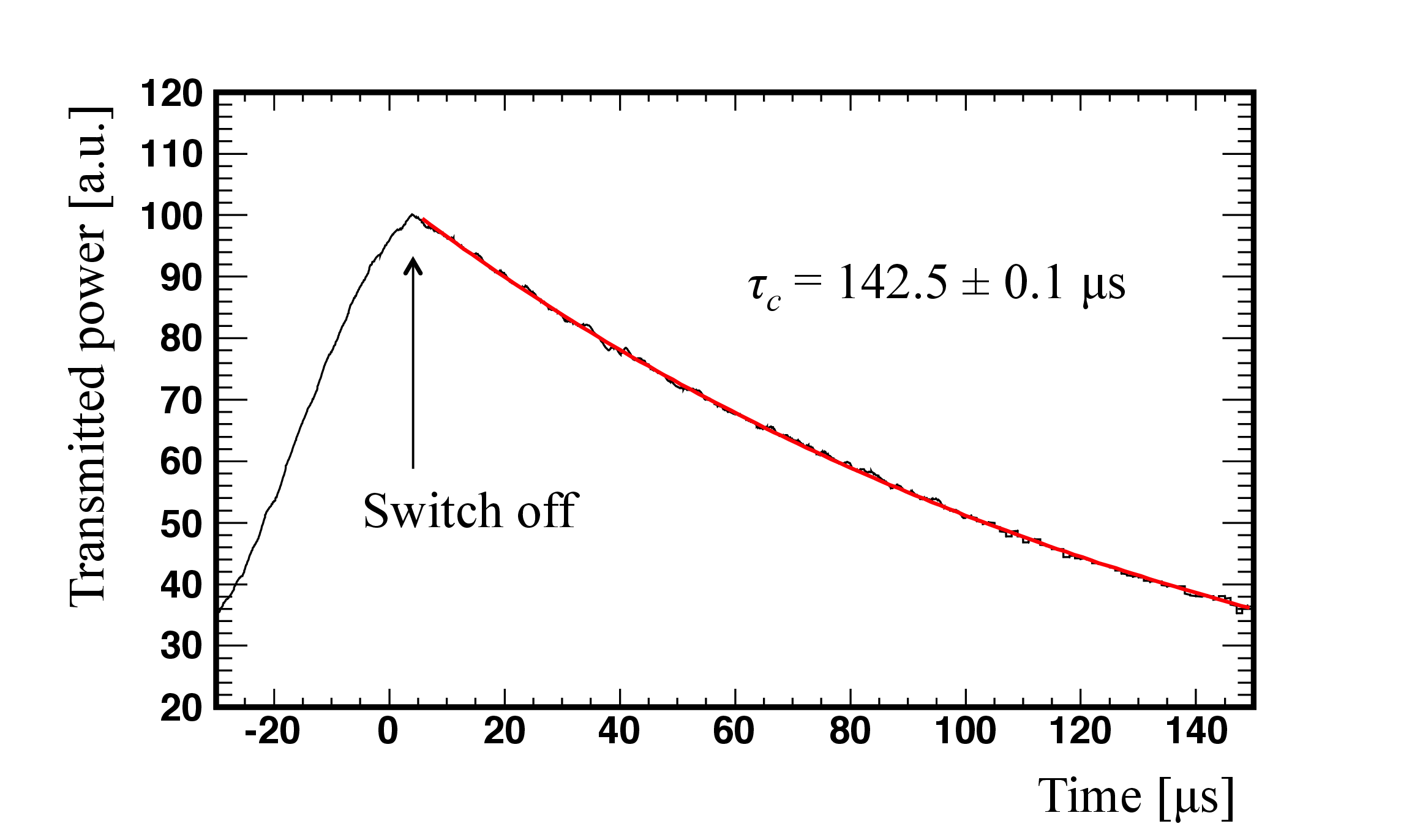}
\caption{The decay of the transmitted power from the cavity. 
The transmitted power was observed while scanning the cavity length, 
and the incident laser light was switched off at its peak.
The red line in the figure is the best fit with the function 
$a e^{-t/\tau_c}+b$ with $a$, $b$, and $\tau_c$ being fitting parameters.  }
\label{fig:decay}
\end{figure}
The decay time was measured to be  $\tau_c=142.5 \pm 0.1 \mu $s
as  shown in figure \ref{fig:decay}.  
Therefore, the finesse $F^{cav}$ of the cavity is calculated to be
\begin{equation}
F^{cav}= 2\pi\tau_c FSR = 646,000 \pm 3,000.\\ 
\label{eqn:cavityfinesse}
\end{equation}
The geometric mean of the reflectance of the two mirrors, $R^{cav} = \sqrt{R_1R_2}$, is estimated 
to be $99.999515 \pm 0.000002  \% $
via the relation 
\begin{equation}
F^{cav} = \frac{\pi \sqrt{R^{cav}}} {1-R^{cav}} 
\label{eqn:cavityproperty}
\end{equation}
The transmittance of M1 and M2 (denoted as $T_1$ and $T_2$, respectively) was also independently 
measured by comparing the laser power incident upon the mirror and transmitted power. 
We obtained the transmittance  of $ T_1 = 4.17 \pm 0.02 $ ppm and $T_2 = 3.77 \pm 0.02 $ ppm, 
where the main source of errors was the linearity of the power measurements.

\section{Results and Discussion}

\subsection{Demonstration of Self-resonation}

\begin{figure}
\includegraphics[width=\linewidth]{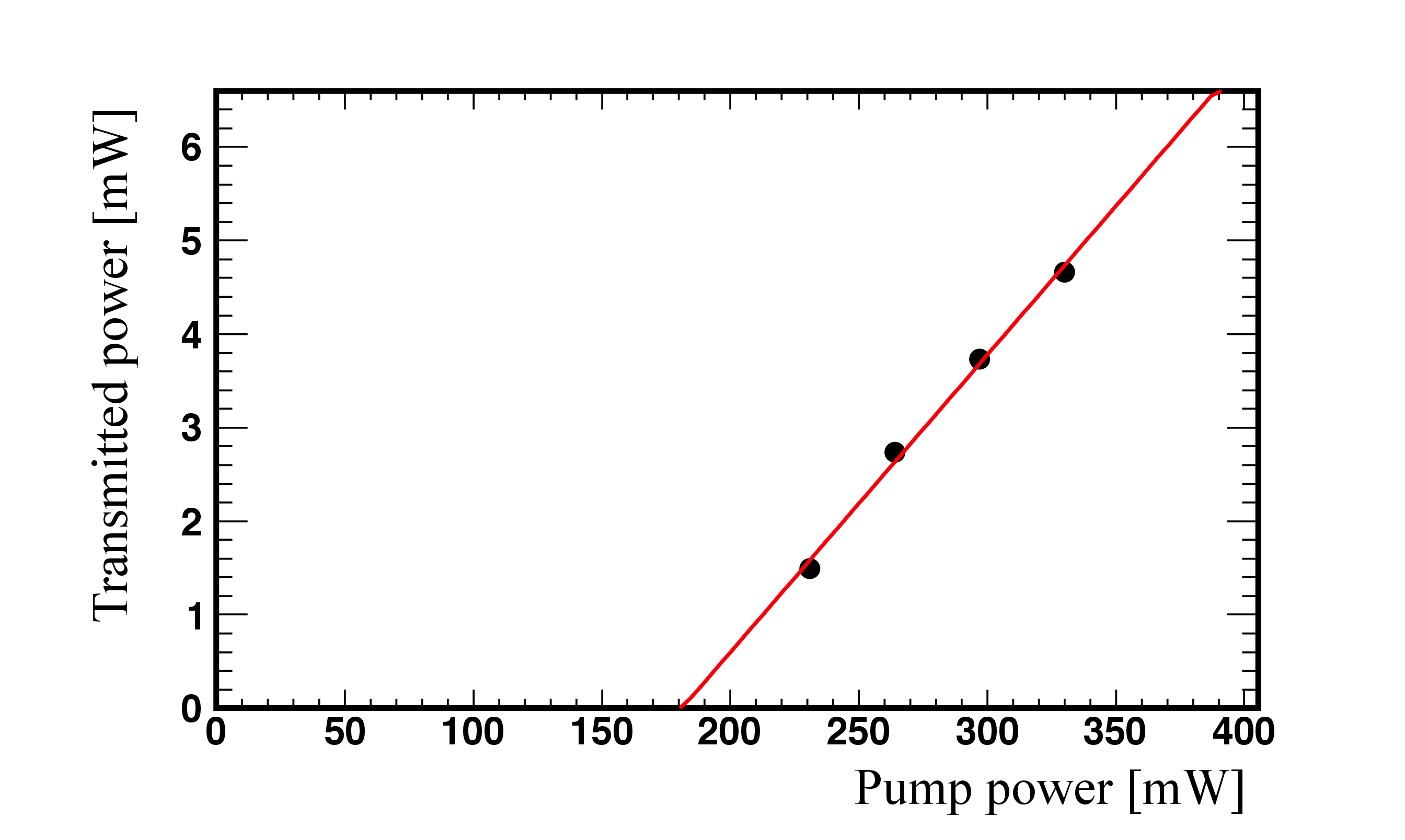}
\caption{The power in the optical loop monitored by PD3. 
The linear dependence on the pump power is observed after the threshold. }
\label{fig:pump}
\end{figure}

\begin{figure}
\includegraphics[width=1.0\linewidth]{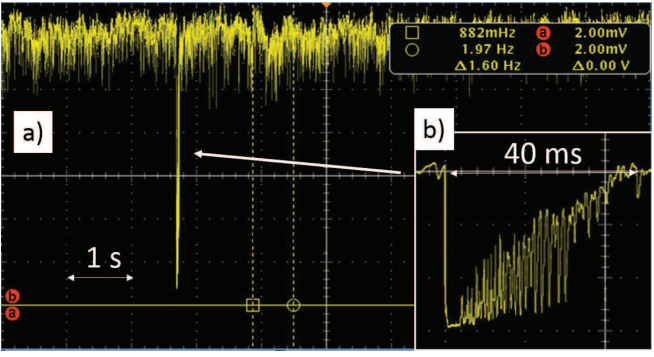}
\caption{The behavior of the self-resonating system monitored by PD3 
when hitting the support tube of the cavity by hand. 
The system went off resonance but resumed by itself a). Figure b) is a magnification 
around the time of the disturbance. The system repeated off- and on-resonant states and resumed to 
a stable condition in approximately 40 ms.} 
\label{fig:hitting}
\end{figure}

After configuring the self-resonating loop,
the power of laser light circulating in the loop, represented by the transmission from M2 monitored by PD3, 
was measured while increasing the driving current of the pump LD (see figure \ref{fig:pump}).
We see that 
the laser power increased linearly with the pump current after a certain threshold,
showing the realization of laser oscillation. 
\begin{figure}
\includegraphics[width=\linewidth]{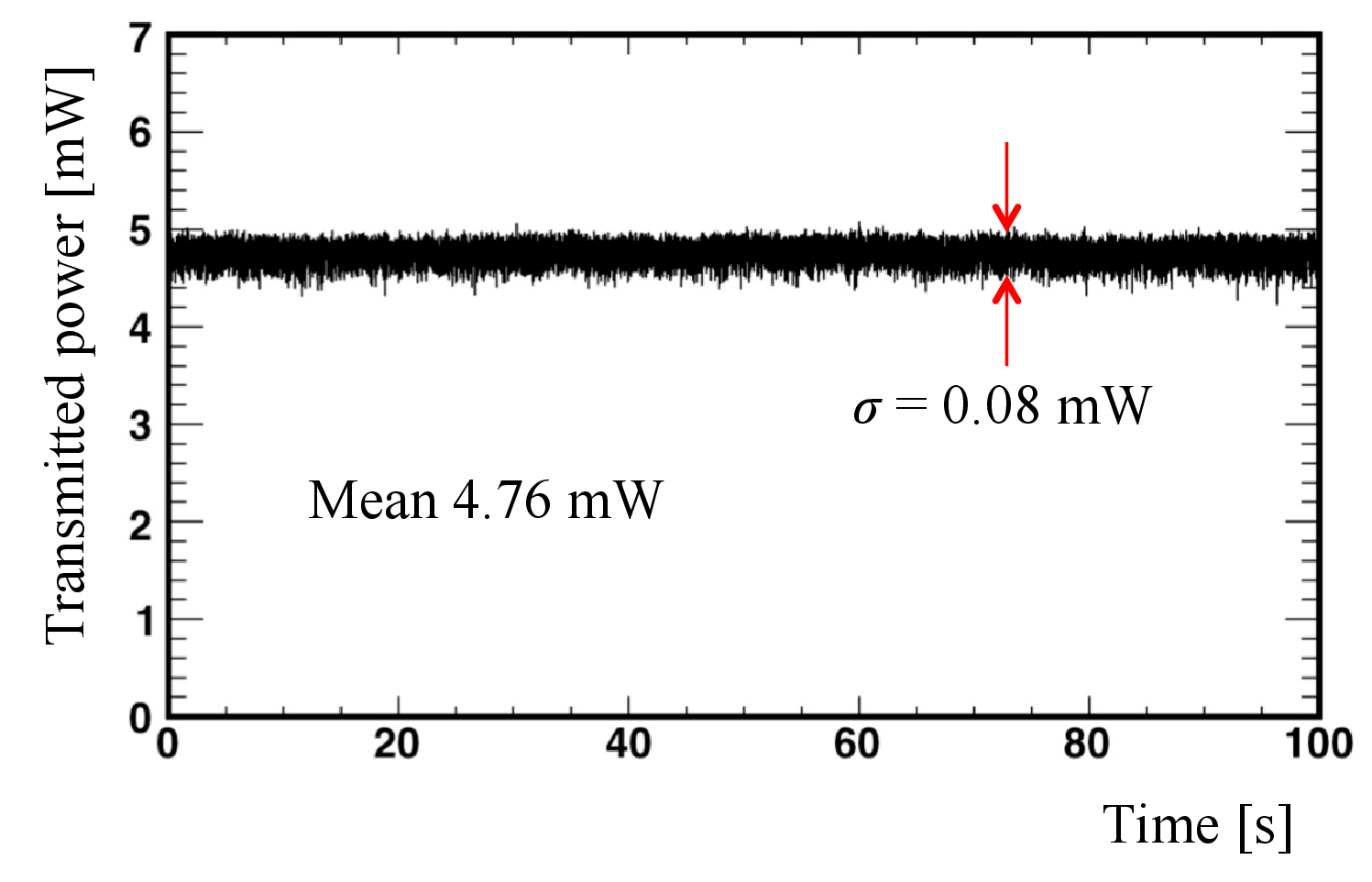}
\caption{The temporal behavior output of PD3 showing stable operation of the self-resonating mechanism.}
\label{fig:temporal}
\end{figure}
In figure \ref{fig:temporal}, 
we show the temporal behavior of the laser power. 
The laser power in the loop was constant
without any deterioration of laser power or its fluctuation.
The fluctuation of the power over the observation period was $1.7 \%$ with
 the average and standard deviation of the fluctuation of   
4.76 mW and 0.08 mW, respectively. 
We demonstrated the long-term operation of the system and found that it was stable for 
more than two hours. 
Figure \ref{fig:hitting} is a demonstration the stability of the system against external disturbances.
When the cavity was in a resonant condition, 
we hit the support tube of the cavity by hand while monitoring 
transmission from M2. 
We saw that the cavity went off-resonance once but resumed by itself.
The time scale of  intensity decrease is approximately 140 $\mu$s, which was
consistent with the lifetime of the laser wave in the cavity.
During the recovery period, 
the cavity repeated off- and on-resonant states for approximately  40 ms, which can be 
considered to be the time scale of the mechanical vibration of the support tube.   

\subsection{Evaluation of the System Performance}
\label{sec:eval}

\begin{figure}
\includegraphics[width=1.0\linewidth]{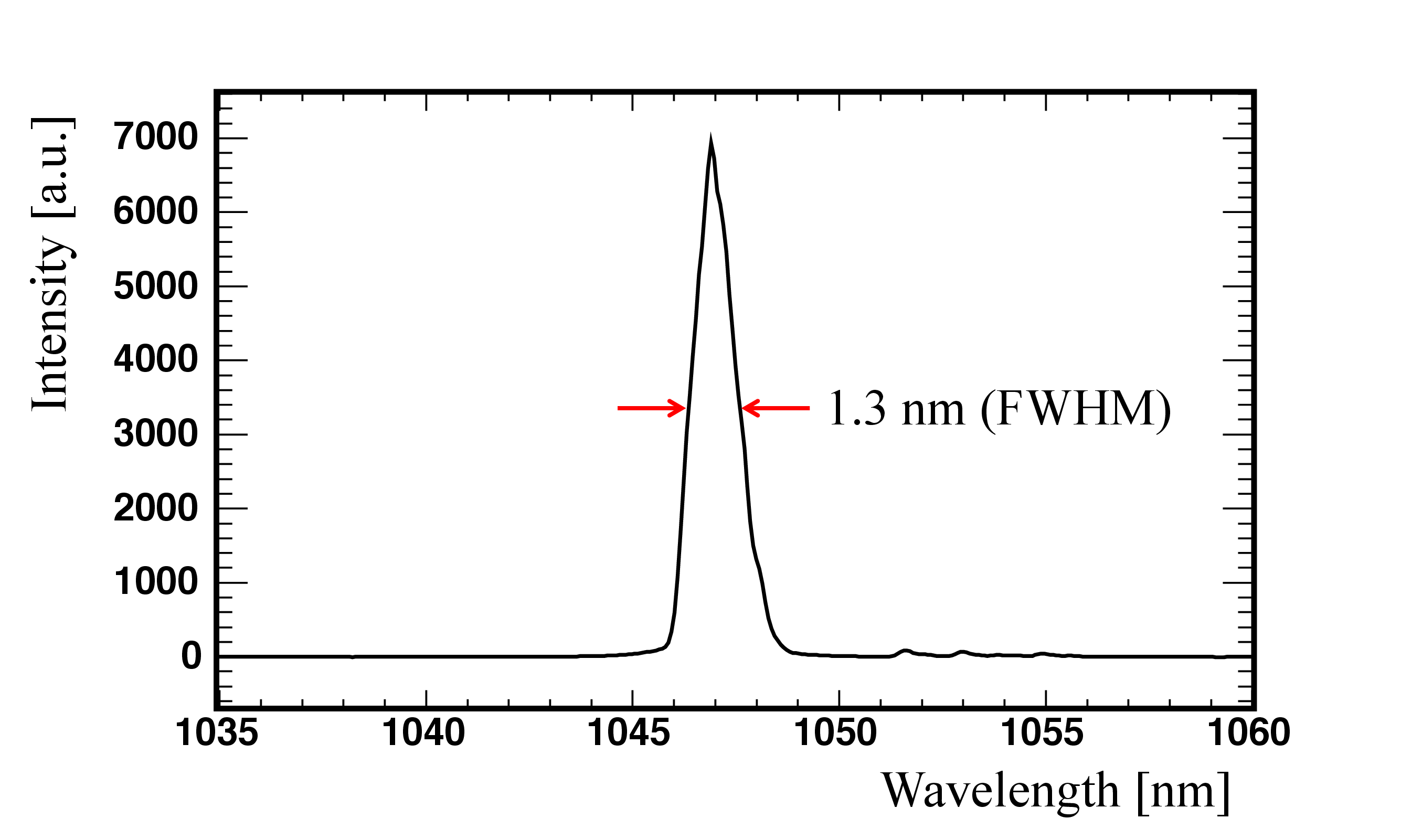}
\caption{The spectrum of laser waves averaged over the oscillation duration. }
\label{fig:spectrum}
\end{figure}

The spectrum of the laser wave taken after Coupler1 is shown in figure \ref{fig:spectrum}.
The spectrum is centered at 1047 nm with a width of 1.3 nm in FWHM. 
The width is much wider than that of the cavity  ($1.5$ kHz),
which we will discuss later.  
We observed a superposition of various oscillations at different frequencies 
over the observation period as a result of self-resonating mechanism.   

\begin{figure}
\includegraphics[width=1.0\linewidth]{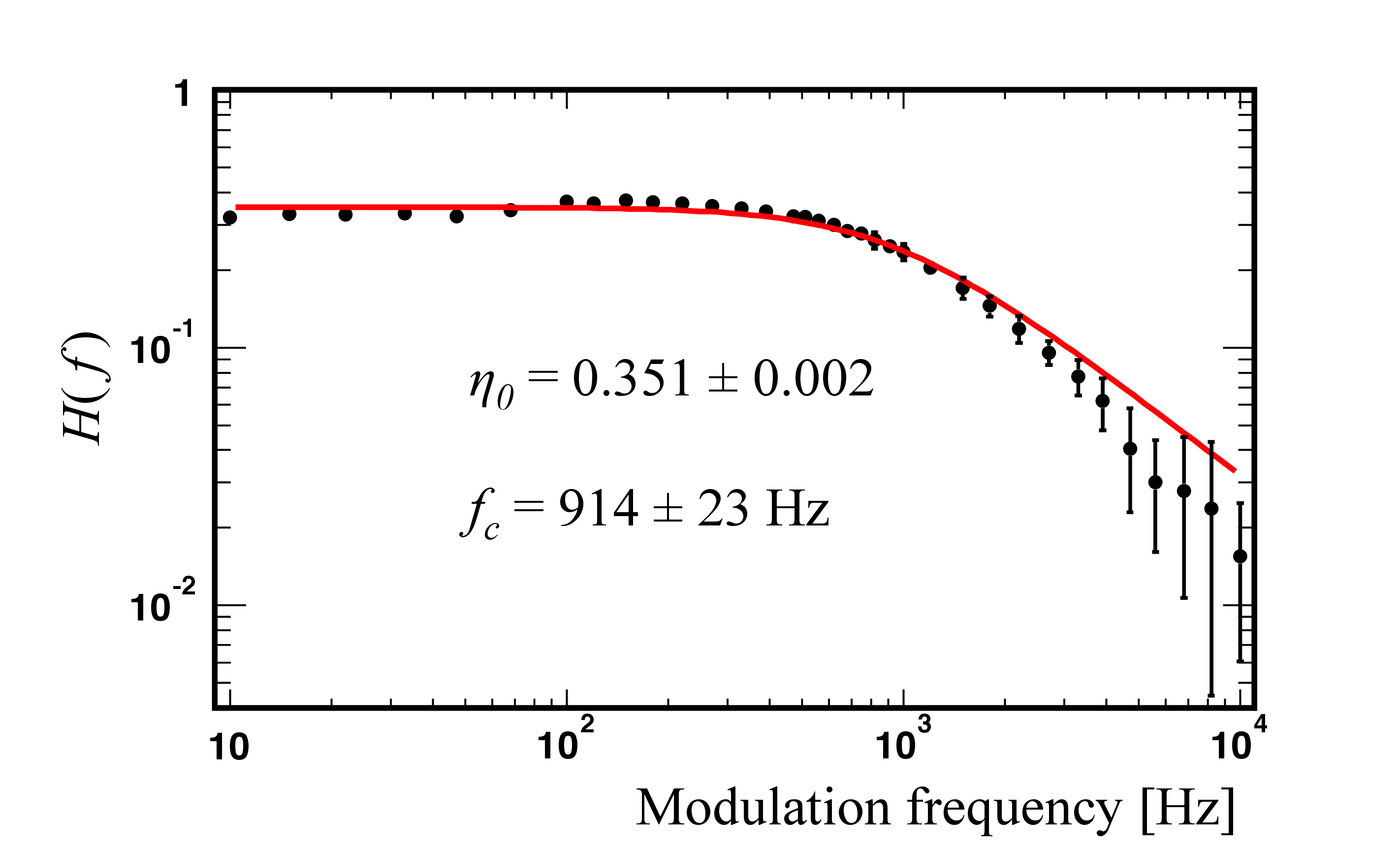}
\caption{The transfer function $H(f)$ with respect to the modulation frequency of pump LD.}
\label{fig:freq1}
\end{figure}

 To quantitatively understand the behavior of the system, 
we measured the laser power circulating in the loop using PD1, PD2, and PD3,
denoted as $P_{PD1}$, $P_{PD2}$ and $P_{PD2}$, respectively.
The measured signal from each detector was calibrated 
to the laser power
using the reflectivity of W1 and W2.   
The accuracy the power measurement is  $5 \%$ and $0.5 \%$ 
for the absolute power and  for its linearity, respectively.
Because the cavity is the most transparent for the resonated laser frequency $f_0$ 
and behaves as a low-pass filter for deviations from $f_0$, 
the performance of the cavity can be evaluated by measuring the transfer function
for laser waves deviating from $f_0$ \cite{NRO}.
To measure the transfer function, 
amplitude modulation was applied on $P_{PD1}$ by modulating 
the power of the pump LD.
The degree of modulation is $1\%$, which is smaller than the power fluctuation 
observed in figure \ref{fig:temporal}  to avoid affecting the 
stability of the laser oscillation.    
The transfer function can be expressed 
by means of 
the modulation frequency, $f$, on $P_{PD1}$
as $H (f)= A_{PD1}(f)/A_{PD3}(f)$,
where $A_{PD1(PD3)}(f)$ is defined as the peak-to-peak amplitude of $P_{PD1(PD3)}(f)$.
 In figure \ref{fig:freq1}, 
$H(f)$ is plotted as a function of  $f$. 
The transfer function is well fitted by the explicit form of the transfer function of the first-order low-pass filter:
\begin{equation}
H (f ) = {\eta _0}/\sqrt {1 + {{\left( {\frac{f }{{f_c }}} \right)}^2}} 
\end{equation}
where  $f_c$  and  $\eta_0$  are fitting parameters.
We obtained $f_c$ and $\eta_0$ of $914 \pm 23$ Hz and $0.351 \pm 0.002$, respectively,
 where $f_c$, and $\eta_0$ can be regarded as the cut-off frequency 
of the filter and transmission efficiency of  the cavity on the resonance, respectively.
The FWHM of the resonance of the cavity is $2fc=2 \times 914 = 1.828 \pm 0.046$ kHz 
because $f_c$ is half of the FWHM. 
 The effective finesse, $F^{eff}$, of the  cavity is calculated to be
\begin{equation}
F^{eff} = \frac{FSR}{2f_c} = 394,000 \pm 10,000 \\
\label{eqn:Floop}
\end{equation}
The obtained finesse is smaller than the intrinsic finesse of $F^{cav}=646,000$  because $F^{eff}$ is a result of the overall
performance the self-resonating mechanism, such as the power stability
or time response of the system against external disturbances, while
$F^{cav}$ represents the intrinsic performance of the cavity determined only  
by the reflectivity of the mirrors.
With the previously measured transmittance of M2 of  $3.77 \pm 0.02 $ ppm 
and the transmitted power of  $P_{out}=4.76 \pm 0.24 $ mW 
(see figure \ref{fig:temporal}), 
the laser power accumulated in the cavity is estimated to be  
$2.52 \pm 0.13 $ kW,
which is approximately  187,000 times more than the incident power  $P_{PD1}$ of 13.5 mW. 

\section{Conclusion and prospect}

We successfully demonstrated the highly stable operation of a feedback-free optical cavity with 
a self-resonating mechanism.
We achieved successful operation of the system with an effective finesse of
 $394,000 \pm 10,000$, and 
the laser power stored in the cavity was estimated to be $2.52 \pm 0.13 $ kW,
showing a power enhancement of approximately  187,000.
In addition, the system 
was highly stable and robust against environmental disturbance, 
having shown
that the power fluctuation was $1.7 \%$, 
and  continuous operation could be achieved for more than two hours.
We emphasize that the required precision for the optical path 
length for this effective finesse is about 2 pm, which had only been achieved
with a sophisticated feedback control system. 
In the current setup, the mirrors of the cavity are mounted on the 
super-invar alloy tube to sustain robustness against environmental 
perturbation. Though the required precision is far higher than the precision
achievable by simply deploying a rigid component, it is still worth 
evaluating the dependence of the system stability on the rigidness of the optical components.

This study showed the possibility of realizing a high finesse cavity 
without a sophisticated active feedback system.
It is highly useful for applications, 
such as  photon facilities by laser-Compton scattering or CEAS, which does not require narrow laser linewidth.
In particular, it is directly applicable to the detection of trace elements absorbers by CEAS.
Since the CEAS obtains information from the change of finesse by trace elements, we can apply the cavity ring-down and/or the response function techniques demonstrated in this paper, which are, in principle, independent of the power stability.

Regarding  high-power-density lasers for laser-Compton facilities, 
it is necessary to develop mode-clocked pulsed oscillation, which we are currently pursuing. 

Issues such as the development of optical components that will be durable for high-power transport, particularly fiber optics,
have yet to be overcome but are expected to be developed in optics communities, 
once the usefulness of this technique is successfully demonstrated.

\begin{acknowledgments}
The authors would like to thank Dr. D. Tatsumi for valuable discussions. 
This work has been supported by JSPS Grants-in-Aid for Scientific Research No. 25246039
and in part by the Quantum Beam Technology Program 
of the Japanese Ministry of
Education, Culture, Sports, Science and Technology. 
\end{acknowledgments}

\end{document}